\documentclass[aps,pre,showpacs,groupedaddress,twocolumn]{revtex4}

\usepackage{amsmath}
\usepackage{amssymb}
\usepackage{graphicx}
\usepackage{color}
\usepackage{bm}
\usepackage{subfigure}
\usepackage{multirow}
\usepackage{latexsym}

\begin{document}
\title{Distribution of randomly diffusing particles in inhomogeneous media}

\author{Yiwei Li}
\author{Osman Kahraman}
\author{Christoph A. Haselwandter}

\affiliation{Department of Physics \& Astronomy and Molecular and Computational Biology Program, Department of Biological Sciences, University of Southern California, Los Angeles, CA 90089, USA}

\begin{abstract}

Diffusion can be conceptualized, at microscopic scales, as the random hopping of particles between neighboring lattice sites. In the case of diffusion in inhomogeneous media, distinct spatial domains in the system may yield distinct particle hopping rates. Starting from the master equations (MEs) governing diffusion in inhomogeneous media we derive here, for arbitrary spatial dimensions, the deterministic lattice equations (DLEs) specifying the average particle number at each lattice site for randomly diffusing particles in inhomogeneous media. We consider the case of free (Fickian) diffusion with no steric constraints on the maximum particle number per lattice site as well as the case of diffusion under steric constraints imposing a maximum particle concentration. We find, for both transient and asymptotic regimes, excellent agreement between the DLEs and kinetic Monte Carlo simulations of the MEs. The DLEs provide a computationally efficient method for predicting the (average) distribution of randomly diffusing particles in inhomogeneous media, with the number of DLEs associated with a given system being independent of the number of particles in the system. From the DLEs we obtain general analytic expressions for the steady-state particle distributions for free diffusion and, in special cases, diffusion under steric constraints in inhomogeneous media. We find that, in the steady state of the system, the average fraction of particles in a given domain is independent of most system properties, such as the arrangement and shape of domains, and only depends on the number of lattice sites in each domain, the particle hopping rates, the number of distinct particle species in the system, and the total number of particles of each particle species in the system. Our results provide general insights into the role of spatially inhomogeneous particle hopping rates in setting the particle distributions in inhomogeneous media.

\end{abstract}

\maketitle

\section{Introduction}
\label{secIntro}

Diffusion processes are of ubiquitous importance throughout science. At
microscopic scales, diffusion may be conceptualized as the random hopping of particles between neighboring lattice sites \cite{Chandrasekhar1943,Berg1993,Codling2008}.
For a given particle species, the particle hopping rate generally depends
on the particular properties of the medium through which the particles diffuse. In inhomogeneous media, particles may therefore show distinct hopping rates in distinct spatial domains of the system. Diffusion in inhomogeneous media occurs in a variety of different contexts, including protein diffusion in cell membranes \cite{Siggia2000,Czondor2012}, ecology \cite{Cantrell1999,Fagan1999,Cronin2003}, earth science \cite{Delay2005,Hoteit2002,LaBolle2000,Zhan2009}, biomedical imaging \cite{Fieremans2010}, and astrophysics \cite{Zhang2000,Marcowith2010}. The general mathematical features of diffusion in inhomogeneous media have been studied extensively \cite{Kampen1987,Kampen1988,Miyazawa1987,Kampen1992} using generalized diffusion equations with spatially-varying diffusion coefficients.
A conceptually and practically important scenario is thereby provided by the diffusion of particles through systems with periodic boundary conditions, in which particles do not directly interact with the system boundaries.

For the case of diffusion in homogeneous media with periodic boundary conditions, the average steady-state distribution of particles is uniform. In contrast, for particles diffusing through inhomogeneous media one generally expects that the average steady-state distribution of particles is non-uniform and dependent on the relative particle hopping rates in the distinct spatial domains of the system. Furthermore, in the pre-asymptotic regime, the average particle distribution in inhomogeneous media may show a complex temporal
evolution towards the steady state of the system. In this article we derive, starting from the master equations (MEs) describing the random hopping of particles in inhomogeneous media, the lattice Langevin equations governing the particle number at each lattice site in the system. We consider the case of free (Fickian) diffusion with no steric constraints on the maximum particle number per lattice site as well as the case of diffusion under steric constraints imposing a maximum particle concentration. The deterministic parts of the lattice Langevin equations provide the deterministic lattice equations (DLEs) specifying the average particle number at each lattice site. From the DLEs we obtain general analytic expressions for the (average) steady-state particle distributions for free diffusion and, in special cases, diffusion under steric constraints in inhomogeneous media. We show that numerical solution of the DLEs offers a computationally efficient method for predicting the (average) distributions of randomly diffusing particles in inhomogeneous media for free diffusion as well as diffusion under steric constraints. For both transient and asymptotic regimes, we test our solutions of the DLEs using kinetic Monte Carlo (KMC) simulations of the underlying MEs. Our results provide general insights into the role of spatially inhomogeneous particle hopping rates in setting the particle distributions in inhomogeneous media. We first consider, in Sec.~\ref{secFreeDiff}, free diffusion in inhomogeneous media. We then consider, in Sec.~\ref{secCrowdDiff}, inhomogeneous systems with steric constraints and single or multiple diffusing particle species. We conclude, in Sec.~\ref{SumCon}, with a summary and discussion of our~results.

\section{Free diffusion}
\label{secFreeDiff}

\begin{figure}[t!]
\includegraphics[width=0.9 \columnwidth]{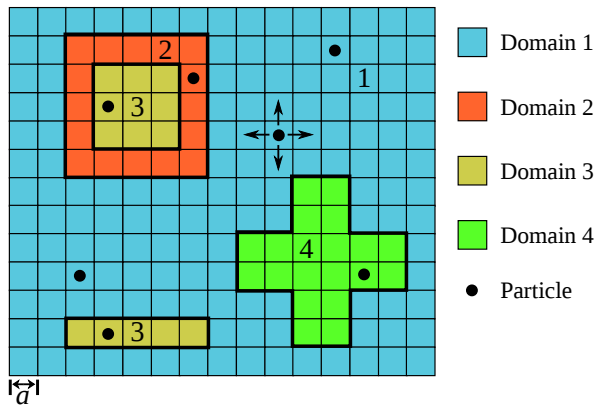} 
\caption{Illustration, for $d=2$, of the hypercubic lattice model of diffusion in inhomogeneous media we consider here. Lattice sites are indicated by unit squares with lattice spacing $a$. A particle occupying a given lattice site is allowed to hop randomly to any one of its $2d$ nearest-neighbor sites. Distinct colors indicate lattice sites with distinct hopping rates $1/\tau_{\alpha(i)}$ for a particle to leave the lattice site, with $\alpha=1,\dots,D$ and $D=4$ here. Domain $\alpha$ encompasses all lattice sites $i$ with hopping rate~$1/\tau_{\alpha(i)}$.}
\label{fig:cartoon}
\end{figure}

We consider in this article particles diffusing in systems with $K$ lattice sites $i=1,2,\dots,K$. We focus on the special case of lattice systems with periodic boundary conditions, but our formalism could be extended to other types of boundary conditions. Throughout this article, we model particle diffusion as the random hopping of particles between nearest-neighbor lattice sites \cite{Chandrasekhar1943,Berg1993,Codling2008}. For simplicity, we focus on hypercubic lattices of dimension $d$ with lattice spacing $a$, implying that each lattice site has $2d$ nearest-neighbor sites (see Fig.~\ref{fig:cartoon}). To model spatially inhomogeneous particle hopping rates, we allow for $D$ distinct domains in the system, labelled by an index $\alpha=1,2,\dots,D$, with the rate for a particle at lattice site $i$ in domain $\alpha$ to hop to a nearest-neighbor lattice site being given by $1/\tau_{\alpha(i)}$. Note, in particular, that if two or more lattice sites have the same hopping rate we consider them to be part of the same domain (Fig.~\ref{fig:cartoon}) irrespective of whether the lattice sites are connected via lattice sites with the same hopping rate, or not. In this section we focus on the case of free (Fickian) diffusion, for which the probability that a given particle hops to a nearest-neighbor lattice site is constant in each domain. In Sec.~\ref{secCrowdDiff} we generalize the formalism developed here to scenarios in which particles interact with each other via steric constraints on the maximum particle number per lattice site and, as a result, the probability for a given particle to hop to a nearest-neighbor lattice site depends on the particle number per lattice site.

\subsection{Stochastic lattice model}

In our stochastic lattice model of diffusion in inhomogeneous media we specify,
at each lattice site $i$, the number of particles (occupation number) through the stochastic variable $N_{i}\geq 0$. We use here the convention that $N_{i}/\epsilon$, with the normalization constant $\epsilon>0$, is the particle number at site $i$. While not necessary for the purposes of the present discussion, using such a normalized $N_{i}$ is convenient \cite{Haselwandter2015,Haselwandter2011,Kahraman2016} if one allows for steric constraints on the occupation number by imposing
a fixed upper limit on $N_{i}$ (see Sec.~\ref{secCrowdDiff}). By definition,
the number of lattice sites in domain $\alpha$, $M_\alpha$, must obey
\begin{equation} \label{eq:vol_conserv}
\sum\limits_{\alpha=1}^D M_{\alpha} = K\,.
\end{equation}
Furthermore, since we use periodic boundary conditions, the total number of particles in the system, $\mathcal{N}$, is conserved:
\begin{equation} \label{eq:prob_conserv}
\frac{1}{\mathcal{N}}\sum\limits_{i=1}^{K} \frac{N_{i}}{\epsilon} = 1 \,.
\end{equation}

The stochastic lattice model we consider here is Markovian and, hence, the state of the system at each time $t$ is completely characterized by the set of occupation numbers (lattice occupancy) ${\bf N}(t)=\{N_{i}(t)\}$ with $1 \leq i \leq K$. The temporal evolution of the lattice occupancy ${\bf N}$ is determined by the ME~\cite{Gardiner1985,Kampen1992}
\begin{equation}
\frac{\partial P}{\partial t}=\sum_{{\bf m}} \!\,\bigl[W({\bf N}-{\bf m};{\bf m}) P({\bf N}-{\bf m},t) - W({\bf N};{\bf m})P({\bf N},t)\bigr]  \,,
\label{df_ME}
\end{equation}
where $P({\bf N},t)$ is the probability that the system is in state ${\bf N}$ at time $t$, $W({\bf N}; {\bf m})$ is the transition rate from lattice
occupancy ${\bf N}$ to lattice occupancy ${\bf N} +{\bf m}$, and ${\bf m}=\{
m_i\}$ with $1 \leq i \leq K$ is the array of jumps in lattice occupancy. For the random hopping of particles to nearest-neighbor ($\text{n.n.}$) sites $j$, the transition rate in Eq.~(\ref{df_ME})
is given by
\begin{align} \label{eq:W_diff}
W({\bf N};{\bf m}) = & \frac{1}{2d\epsilon}\sum\limits_i \frac{N_{i}}{\tau_{\alpha(i)}}
\delta(m_i+\epsilon)   \nonumber \\
& \times \sum\limits_{j \text{~n.n.~of~}i} \delta(m_{j}-\epsilon) \prod_{k\neq i,j} \delta(m_k) \,,
\end{align}
where $2d$ is the coordination number of the hypercubic lattice in $d$ dimensions,
the factor of $1/\epsilon$ arises because we use the convention that $1/\tau_{\alpha(i)}$ is the hopping rate per particle, and $\delta(x)$ is the Dirac-delta function. We use Dirac-delta functions, rather than Kronecker-delta functions, in Eq.~(\ref{eq:W_diff}) in order to make the connection between the ME~(\ref{df_ME}) and the corresponding DLEs in Eq.~(\ref{eq:df_DLLE}) more transparent (see below), which amounts to replacing the summation in the ME~(\ref{df_ME}) by an integral over all (continuous) ${\bf m}$ \cite{Haselwandter2007,Li2017}.

\subsection{Deterministic lattice equations}
\label{secDLEfree}

Following the approach in Refs.~\cite{Haselwandter2015,Kampen1992,Haselwandter2007,Li2017,Fox1991,Horsthemke1977}
we transform the ME~(\ref{df_ME}) into the more tractable lattice Langevin equations
\begin{align} \label{eq:LE}
 \frac{d N_{i}}{d t} = K_i^{(1)} + \eta_i \,,
\end{align}
where the $\eta_i$ are Gaussian noises with zero mean and covariance
\begin{align} \label{eq:noise_cov}
 \langle \eta_i(t_1)\eta_j(t_2) \rangle = K_{i,j}^{(2)} \delta(t_1-t_2) \,,
\end{align}
the $K_i^{(1)}$ and $K_{i,j}^{(2)}$ are the first and second moments of the transition rate in Eq.~(\ref{eq:W_diff}),
\begin{eqnarray} \label{eq:1st_moment}
 && K_i^{(1)}({\bf N}) = \int m_i W({\bf N}; {\bf m}) d{\bf m} \,, \\ \label{eq:2nd_moment}
 && K_{i,j}^{(2)}({\bf N}) = \int m_i m_j W({\bf N}; {\bf m}) d{\bf m} \,,
\end{eqnarray}
and we have taken the $N_{i}$ to be continuous variables 
\cite{Haselwandter2015,Kampen1992,Haselwandter2007,Li2017,Fox1991,Horsthemke1977}. The deterministic parts of the lattice Langevin equations~(\ref{eq:LE}) yield the DLEs associated with the ME~(\ref{df_ME}) with Eq.~(\ref{eq:W_diff}):
\begin{align} \label{eq:df_DLLE}
\frac{d \phi_{i}}{d t} = -\frac{\phi_{i}}{\tau_{\alpha(i)}} + \frac{1}{2d}\sum_{j\text{~n.n.~of~}i} \frac{\phi_{j}}{\tau_{\alpha(j)}} \,,
\end{align}
where the $\phi_{i}$ are the average $N_{i}$, $\phi_{i}(t)=\langle N_{i}(t) \rangle$. The first (negative) term in Eq.~(\ref{eq:df_DLLE}) arises
from the average rate for particles to hop away from site $i$, while the other (positive) terms in Eq.~(\ref{eq:df_DLLE}) correspond to the hopping of particles from the nearest-neighbor sites of site $i$ to site $i$. Since the noise in Eq.~(\ref{eq:LE}) has zero mean, Eq.~(\ref{eq:prob_conserv}) implies that $\phi_{i}$ is conserved:
\begin{align} \label{eq:con_fd}
\frac{1}{\mathcal{N}} \sum \limits_{i=1}^K \frac{\phi_{i}}{\epsilon} = 1\,.
\end{align}
For a given set of initial conditions $\{\phi_i(0)\}$, the DLEs~(\ref{eq:df_DLLE})
can be readily solved numerically using standard methods \cite{Mathematica}, yielding a unique solution for the average particle occupancies $\phi_i(t)$ for all $i$ and all $t$. Indeed, in the steady state of the system with $d
\phi_i/dt=0$ for all $i$, the DLEs in Eq.~(\ref{eq:df_DLLE}) together with the constraint in Eq.~(\ref{eq:con_fd}) fixing the total particle number in the system constitute a set of $K$ linearly independent algebraic equations, which uniquely specify the steady-state $\phi_{i}$ at each lattice site.

\subsection{Particle distribution}
\label{secDLEfreePD}

We characterize the (average) distribution of randomly diffusing particles in inhomogeneous media through the average fraction of all particles in domain $\alpha$, $F_\alpha$. In terms of the solutions of the DLEs~(\ref{eq:df_DLLE}), $F_\alpha$ can be expressed as
\begin{equation} \label{Falphadef}
F_\alpha=\frac{\sum_{i\textrm{~in~domain~}\alpha}\phi_{i}}{\sum_{i=1}^K
\phi_{i}}\,.
\end{equation}
As described in Sec.~\ref{secDLEfree}, the $\phi_i(t)$ are readily obtained
numerically from the DLEs~(\ref{eq:df_DLLE}) with Eq.~(\ref{eq:con_fd}),
from which $F_\alpha$ can be computed by directly evaluating Eq.~(\ref{Falphadef}). At least for special cases, it is also feasible to obtain analytic expressions of $F^\alpha$. In particular, we construct the steady-state (s.s.) particle distribution $F_\alpha^\textrm{(s.s.)}$ by setting the left-hand side of Eq.~(\ref{eq:df_DLLE}) equal to zero, and matching positive and negative terms on the right-hand side of Eq.~(\ref{eq:df_DLLE}). Note that Eq.~(\ref{eq:df_DLLE}) then implies that, in the steady state of the system, all the $\phi_{i}$ lying in a particular domain $\alpha$ with hopping rate $1/\tau_\alpha$ take the same value $\phi^{(\alpha)}$. More generally, Eq.~(\ref{eq:df_DLLE}) implies that, in the steady state of the system, the particle occupancies in any two domains $\alpha$ and $\beta$ satisfy
\begin{equation} \label{eqssrelationFD}
\frac{\phi^{(\alpha)}}{\tau_\alpha} = \frac{\phi^{(\beta)}}{\tau_\beta}\,,
\end{equation}
as also expected based on the principle of detailed balance. We thus find that, in the steady state of the system, the average fraction of all particles in domain $\alpha$ is given~by
\begin{equation} \label{eq:df_sol2}
F_{\alpha}^\textrm{(s.s.)} = \frac{M_\alpha \phi^{(\alpha)}}{\sum\limits_{\beta=1}^D M_{\beta}\phi^{(\beta)}}
= \frac{M_{\alpha}\tau_{\alpha}}{\sum\limits_{\beta=1}^D M_{\beta}\tau_{\beta}}\,,
\end{equation}
where $M_{\alpha}\tau_{\alpha}$ corresponds to the characteristic time a randomly hopping particle spends in domain $\alpha$. Thus, the steady-state particle fraction in domain $\alpha$ is directly proportional to the inverse of the hopping rate in domain $\alpha$, and to the number of lattice sites in domain $\alpha$. Note, in particular, that $F_{\alpha}^\textrm{(s.s.)}$ is independent of the system geometry, i.e., the arrangement and shape of domains, as well as the system dimensionality $d$. Since we do not allow
here for any interactions between particles, the results in Eqs.~(\ref{eq:df_DLLE}),
(\ref{eqssrelationFD}), and (\ref{eq:df_sol2}) readily generalize to an arbitrary number of different (non-interacting) particle species.

Some further insight into the steady-state distribution of randomly diffusing
particles in inhomogeneous media can be gained by drawing an analogy between
Eq.~(\ref{eqssrelationFD}) and the self-assembly of particle aggregates in dilute solutions \cite{BenShaul1994,Safran2003}. In particular, introducing a constant $\mu$, Eq.~(\ref{eqssrelationFD}) can be rewritten~as
\begin{align} \label{eq:thermo_free1}
\epsilon_\alpha + \log \phi^{(\alpha)} = \mu
\end{align}
for any domain $\alpha$, where $\epsilon_\alpha=\log \left(\tau_0/\tau_\alpha\right)$, in which $\tau_0$ is a constant. Viewed as an equation for $\phi^{(\alpha)}$, Eq.~(\ref{eq:thermo_free1}) takes the same basic form as the thermodynamic equilibrium distribution of self-assembled particle aggregates in dilute solutions with energy $\epsilon_\alpha$ per particle in particle aggregate
$\alpha$ and particle chemical potential $\mu$ \cite{BenShaul1994,Safran2003}. From Eq.~(\ref{eq:thermo_free1}), together with the constraint $\sum_\alpha M_\alpha \phi^{(\alpha)} =\mathcal{N} \epsilon$ implied by Eq.~(\ref{eq:con_fd}), we find \begin{align} \label{chemPotFD} e^{\mu} = \frac{\mathcal{N}\epsilon}{\sum\limits_{\alpha=1}^D
M_\alpha e^{-\epsilon_\alpha}}\,.
\end{align}
Equations~(\ref{eq:thermo_free1}) and~(\ref{chemPotFD}) allow us to construct a general expression for the steady-state lattice occupancies for freely diffusing particles in inhomogeneous media,
\begin{align} \label{eq:thermo_free4}
\phi^{(\alpha)} 
 = \frac{\mathcal{N}\epsilon \tau_\alpha}{\sum\limits_{\beta=1}^D M_\beta \tau_\beta}
 \, ,
\end{align}
which yields the same expression for $F_{\alpha}^\textrm{(s.s.)}$ as in Eq.~(\ref{eq:df_sol2}).

\subsection{Simulation of free diffusion}
\label{secSimFD}

\begin{figure}[t!]
\includegraphics[width=\columnwidth]{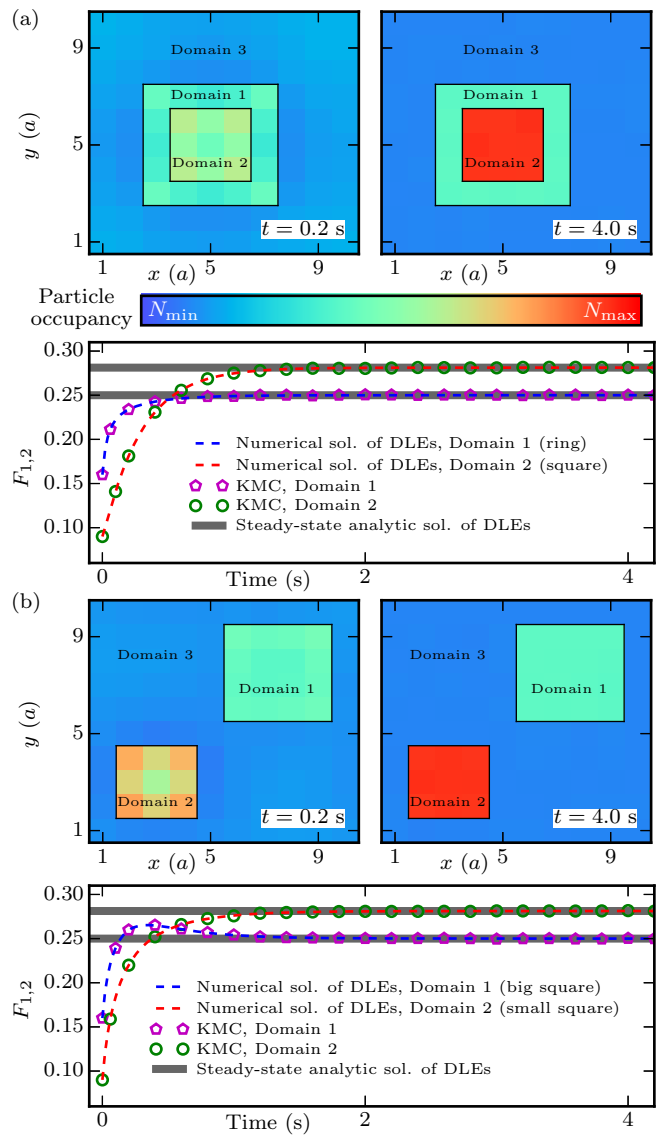}
\caption{Free diffusion of particles for $d=2$ and three domains with $1/\tau_1=32$~s$^{-1}$, $1/\tau_2=16$~s$^{-1}$, and $1/\tau_3=80$~s$^{-1}$ employing the two distinct system geometries shown in (a) and (b). We used $K=100$ with $M_1=16$ and $M_2=9$, periodic boundary conditions, and a homogeneous initial particle distribution $N_i(0)=30\epsilon$ with $\epsilon=1/100$. The upper panels
in (a,b) show $\langle N_i \rangle$ in transient and steady-state regimes obtained from KMC simulations of the ME~(\ref{df_ME}) with Eq.~(\ref{eq:W_diff}). The minima and maxima of the plotted $\langle N_i \rangle$ are $(N_{\rm{min}},N_{\rm{max}})=(0.1,1.0)$,
and we use the same color bar in (b) as in (a). The lower panels in (a,b) show the temporal evolution of $F_{1,2}$. The dashed curves indicate
numerical solutions of the DLEs~(\ref{eq:df_DLLE}), the symbols denote averages over KMC simulations of the ME~(\ref{df_ME}) with Eq.~(\ref{eq:W_diff}), and the gray lines show the steady-state analytic solutions in Eq.~(\ref{eq:df_sol2}).
All KMC results were averaged over 1000 independent realizations each.}
\label{fig:3doman_1type_free}
\end{figure}

As discussed in Secs.~\ref{secDLEfree} and~\ref{secDLEfreePD}, the DLEs~(\ref{eq:df_DLLE})
allow prediction of the (average) transient and steady-state distributions of particles diffusing freely through inhomogeneous media. To test these predictions, we carried out KMC simulations of the ME~(\ref{df_ME}) with Eq.~(\ref{eq:W_diff}). For our KMC simulations we used the Next Subvolume Method \cite{Elf2003}. In particular, we considered 2D systems with three distinct domains (see Fig.~\ref{fig:3doman_1type_free}). Keeping the values of $M_{1,2,3}$ fixed, we allowed for two distinct system geometries. On the
one hand, we considered a scenario in which one domain, with the shape of
a square, was enclosed by the other two domains [see Fig.~\ref{fig:3doman_1type_free}(a)]. On the other hand, we considered a system geometry with two separate square-shaped
domains enclosed by a third domain [see Fig.~\ref{fig:3doman_1type_free}(b)]. For both of these two system geometries, we find excellent agreement between the $F_\alpha$ obtained from the DLEs~(\ref{eq:df_DLLE}), the $F_\alpha^\text{(s.s.)}$ obtained from Eq.~(\ref{eq:df_sol2}), and the corresponding $F_\alpha$ obtained by averaging over KMC simulations of the ME~(\ref{df_ME}) with Eq.~(\ref{eq:W_diff}) (Fig.~\ref{fig:3doman_1type_free}).

As predicted by the steady-state analytic solutions in Eq.~(\ref{eq:df_sol2}),
we find that the steady-state particle distributions obtained from the DLEs~(\ref{eq:df_DLLE}) and KMC simulations of the ME~(\ref{df_ME}) with Eq.~(\ref{eq:W_diff}) are spatially inhomogeneous with $\phi^{(\alpha)} \propto \tau_\alpha$. Furthermore,
as predicted by Eq.~(\ref{eq:df_sol2}), the $F_\alpha^\textrm{(s.s.)}$ obtained from the DLEs~(\ref{eq:df_DLLE}) and KMC simulations of the ME~(\ref{df_ME}) with Eq.~(\ref{eq:W_diff}) are independent of the system geometry considered, with $F_\alpha^\textrm{(s.s.)} \propto M_\alpha \tau_\alpha$. In contrast, the temporal evolution of the particle distribution strongly depends on the system geometry, in both the ME~(\ref{df_ME}) with Eq.~(\ref{eq:W_diff})
and the DLEs~(\ref{eq:df_DLLE}). We find that, in the case of two square-like domains with $\tau_3 < \tau_1 < \tau_2$, domain 1 shows a pronounced ``overshoot'' in $F_1$ [Fig.~\ref{fig:3doman_1type_free}(b)]. No such overshoot is obtained in Fig.~\ref{fig:3doman_1type_free}(a). We attribute the observed overshoot in $F_1$ in Fig.~\ref{fig:3doman_1type_free}(b) to a slow equilibration between domains 1 and 2 in Fig.~\ref{fig:3doman_1type_free}(b). In agreement with this picture, we find that the magnitude of the overshoot in Fig.~\ref{fig:3doman_1type_free}(b) decreases if the distance between domains 1 and 2 is reduced.

\section{Diffusion under steric constraints}
\label{secCrowdDiff}

In this section we generalize the formalism developed in Sec.~\ref{secFreeDiff} to allow for interactions between randomly diffusing particles in crowded
environments in the form of steric constraints. In particular, we impose the constraint that, at each lattice site $i$, the particle occupancy cannot increase beyond $N_{i}=1$, which means that $N_{i}$ is restricted to the range $0\leq N_{i} \leq 1$, with the maximum particle number per lattice site being given by $1/\epsilon$. To implement particle crowding in our stochastic lattice model we use a phenomenological approach, and assume that the rates of all diffusion processes increasing the particle number at lattice site $i$ are $\propto\left[1 - f_i\left(N_{i}\right)\right]$, where $0\leq f_i(N_{i}) \leq 1$. The form of the function $f_i(x)$ will, in general, depend on the particular properties of the system under consideration. For instance, if the steric constraints in the system are non-uniform, different $f_i(x)$ may need to be used for different lattice sites. We focus here on the most straightforward choice of a uniform $f_i(x)=x$ that has previously been successfully employed in the context of population biology \cite{Satulovsky1996,McKane2004,Lugo2008}, protein diffusion in crowded cell membranes \cite{Kahraman2016,Li2017,Haselwandter2015,Haselwandter2011}, and general models of non-Fickian diffusion \cite{Fanelli2010,Fanelli2013}. We first consider, in Secs.~\ref{subsec:one-type} and~\ref{secSimSingleSteric}, the case of a single particle species diffusing through inhomogeneous media under steric constraints and then, in Secs.~\ref{subsec:multi-type} and~\ref{secSimMulti}, allow for multiple diffusing particle species.

\subsection{Single particle species}
\label{subsec:one-type}

As in Sec.~\ref{secFreeDiff}, our stochastic lattice model of particles diffusing
through inhomogeneous media under steric constraints is defined by the ME~(\ref{df_ME}). However, the transition rate in the ME~(\ref{df_ME}) now takes the form
\begin{align} \label{eq:W_diff2}
W({\bf N};{\bf m}) =& \frac{1}{2d\epsilon}\sum\limits_i \frac{N_{i}}{\tau_{\alpha(i)}}
\delta(m_i+\epsilon)  \nonumber \\ & \times \sum\limits_{j\text{~n.n.~of~}i} (1-N_{j})\delta(m_{j}-\epsilon) \prod_{k\neq i,j} \delta(m_k) \,.
\end{align}
Proceeding as in Sec.~\ref{secFreeDiff}, we find that the DLEs associated
with the ME~(\ref{df_ME}) with Eq.~(\ref{eq:W_diff2}) are given by 
\begin{equation} \label{eq:cd_DLLE}
\frac{d \phi_{i}}{d t} = -\frac{\phi_{i}}{2d\tau_{\alpha(i)}} \sum_{j\text{~n.n.~of~}i}(1-\phi_{j})
+\frac{1-\phi_{i}}{2d}\sum_{j\text{~n.n.~of~}i}\frac{\phi_{j}}{\tau_{\alpha(j)}} \,,
\end{equation}
where, as in Sec.~\ref{secFreeDiff}, the negative (positive) terms correspond to the hopping of particles away from (to) lattice site $i$. The DLEs~(\ref{eq:cd_DLLE}) are readily solved numerically starting from a given set of initial conditions $\{\phi_i(0)\}$ using standard methods \cite{Mathematica}, which uniquely specifies $\phi_i(t)$ and hence allows computation of $F_\alpha$ through direct evaluation of Eq.~(\ref{Falphadef}).

The steady-state particle distribution $F_\alpha^{\textrm{(s.s.)}}$ can be calculated following similar steps as in Sec.~\ref{secFreeDiff}. We first note that, as in Sec.~\ref{secFreeDiff}, Eq.~(\ref{eq:cd_DLLE}) suggests that, in the steady state of the system, all the $\phi_{i}$ lying in a particular domain $\alpha$ with hopping rate $1/\tau_\alpha$ take the same value $\phi^{(\alpha)}$. Equation~(\ref{eqssrelationFD}) now generalizes to
\begin{align} \label{eq:cd_relation}
\frac{1}{1-\phi^{(\alpha)}} \frac{\phi^{(\alpha)}}{\tau_\alpha} = \frac{1}{1-\phi^{(\beta)}} \frac{\phi^{(\beta)}}{\tau_\beta}\,.
\end{align}
From the above relations, together with Eq.~(\ref{eq:con_fd}), the average
steady-state occupancies $\phi^{(\alpha)}$ are readily obtained numerically \cite{Mathematica}, from which we compute $F_\alpha^{\textrm{(s.s.)}}$ by evaluating Eq.~(\ref{Falphadef}). At least for special cases, it is also
feasible to analytically solve Eq.~(\ref{eq:cd_relation}) with Eq.~(\ref{eq:con_fd})
for the steady-state particle distribution. For instance, consider a system with two domains $\alpha=1,2$. In this case, Eqs.~(\ref{eq:con_fd}) and~(\ref{eq:cd_relation}) yield
\begin{equation} \label{solphi2D2}
\phi^{(2)}=\frac{\tau_2 \phi^{(1)}}{\left(\tau_2-\tau_1\right) \phi^{(1)}+\tau_1}\,,
\end{equation}
with $\phi^{(1)}$ fixed by the quadratic equation
\begin{align} \label{phi1SCEx}
\mathcal{A} \left(\phi^{(1)}\right)^2 + \mathcal{B} \phi^{(1)} + \mathcal{C} = 0 \,,
\end{align}
where $\mathcal{A}=M_1(\tau_2-\tau_1)$, $\mathcal{B}=M_1 \tau_1 + M_2 \tau_2 - \mathcal{N}\epsilon \left(\tau_1-\tau_2\right)$, and $\mathcal{C}=-\mathcal{N}\epsilon
\tau_1$. Equation~(\ref{phi1SCEx}) admits the two solutions
\begin{align}
\phi^{(1)}_\pm = \frac{-\mathcal{B}\pm \left(\mathcal{B}^2-4\mathcal{A}\mathcal{C}\right)^{1/2}}{2\mathcal{A}} \,.
\end{align}
Since $0\leq \phi^{(\alpha)} \leq 1$, $\phi^{(1)}=\phi^{(1)}_+$ is the only physically relevant solution (see below), which yields $\phi^{(2)}$ via Eq.~(\ref{solphi2D2}),
and hence $F_{1,2}^{\textrm{(s.s.)}}$ through Eq.~(\ref{Falphadef}):
\begin{equation}
F_{1,2}^{\textrm{(s.s.)}} = \frac{M_{1,2} \phi^{\textrm{(1,2)}}}{M_{1} \phi^{\textrm{(1)}}+M_{2} \phi^{\textrm{(2)}}} \,.
\end{equation}
The above analytic solution procedure can be generalized to more complicated systems with $D>2$, which generally requires solution of a $D$-th order polynomial. Note that, as for the case of free diffusion, the $F_{\alpha}^\textrm{(s.s.)}$ for diffusion under steric constraints implied by Eq.~(\ref{eq:cd_relation}) with Eq.~(\ref{eq:con_fd}) are independent of the arrangement and shape of domains, as well as the system dimensionality.

As in Sec.~\ref{secFreeDiff}, the DLEs~(\ref{eq:cd_DLLE}) are expected to yield a single (unique) physically relevant steady-state solution. To see this explicitly, it is convenient to rewrite Eqs.~(\ref{eq:con_fd}) and~(\ref{eq:cd_relation}) in the form
\begin{eqnarray} \label{eq:example2_b} \sum_{\alpha=1}^D M_\alpha \phi^{(\alpha)} &=& \mathcal{N}\epsilon\,,\\ \label{eq:example2_a}
\left[1-f\left(\phi^{(\beta)}\right)\right] \frac{\phi^{(\alpha)}}{\tau_\alpha} &=& \left[1-f\left(\phi^{(\alpha)}\right)\right] \frac{\phi^{(\beta)}}{\tau_\beta} \,,
\end{eqnarray}
where we have allowed for a generalized steric constraint $\propto (1-f(N_i))$, with $f(x)$ being a monotonically increasing function of $x$ and $0\leqslant f(x)\leqslant 1$. Assume that, as in the example of a system with two domains
considered above, domain 1 admits two solutions $\phi^{(1)}_\pm$, and let
$\phi^{(1)}_+ > \phi^{(1)}_-$. If the system is initially in a steady state with $\phi^{(1)}=\phi^{(1)}_+$ then, according to Eq.~(\ref{eq:example2_b}), a transition to a competing steady state with $\phi^{(1)}=\phi^{(1)}_-$ would require an increase in the value of at least one $\phi^{(\gamma)}$ with $\gamma\neq 1$. But a decrease in $\phi^{(\alpha)}$ produces an \textit{increase} in $\left[1-f\left(\phi^{(\alpha)}\right)\right]$
if $0 \leq \phi^{(\alpha)} \leq 1$, and vice versa, resulting in violation of Eq.~(\ref{eq:example2_a}). Thus, the physically relevant steady-state
solutions $\phi^{(\alpha)}$ for particles diffusing through inhomogeneous media under steric constraints are expected to be unique with, starting from a given set of initial conditions $\{\phi_i(0)\}$, $\phi_i(t)$ being uniquely determined by the DLEs~(\ref{eq:cd_DLLE}) for all $i$ and all $t$. Finally, we note that Eq.~(\ref{eq:cd_relation}) with Eq.~(\ref{eq:con_fd}) can be connected to the thermodynamic formalism describing the self-assembly of particle aggregates in dilute solutions \cite{BenShaul1994,Safran2003} following similar steps as in Sec.~\ref{secDLEfreePD}. We return to this point in Sec.~\ref{subsec:multi-type}.

\subsection{Simulation of single-species diffusion under steric constraints}
\label{secSimSingleSteric}

\begin{figure}[t!]
\includegraphics[width=\columnwidth]{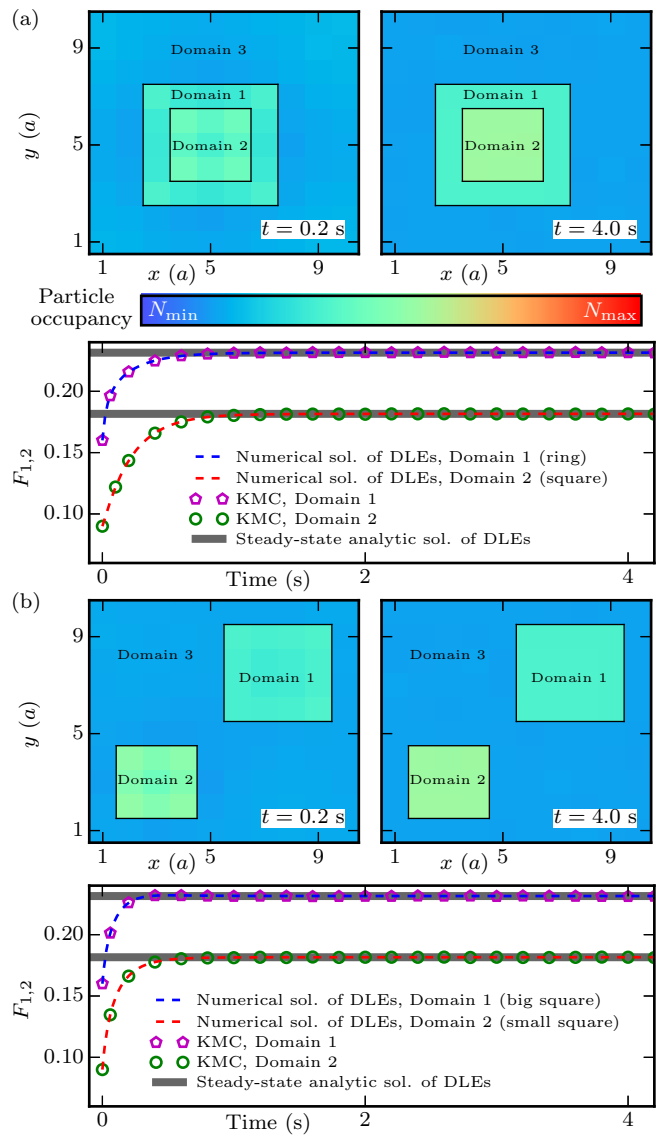}
\caption{Same results as in Fig.~\ref{fig:3doman_1type_free}, but for a single particle species diffusing through inhomogeneous media under steric constraints, described by the ME~(\ref{df_ME}) with Eq.~(\ref{eq:W_diff2}), the DLEs~(\ref{eq:cd_relation}),
and the analytic solutions of the steady-state relations in Eq.~(\ref{eq:cd_relation}) with Eq.~(\ref{eq:con_fd}), for the two distinct system geometries shown
in (a) and (b). We use the same color bar as in Fig.~\ref{fig:3doman_1type_free}.}
\label{fig:3doman_1type_crowd}
\end{figure}

As in Sec.~\ref{secSimFD}, we tested the accuracy of the particle distributions predicted by the DLEs~(\ref{eq:cd_DLLE}), with the steady-state
particle distributions implied by Eq.~(\ref{eq:cd_relation}) with Eq.~(\ref{eq:con_fd}),
by carrying out KMC simulations of the ME~(\ref{df_ME}) with Eq.~(\ref{eq:W_diff2})
using the Next Subvolume Method \cite{Elf2003}. We first considered the same system geometries and parameter values as in Fig.~\ref{fig:3doman_1type_free},
but for diffusion under steric constraints (see Fig.~\ref{fig:3doman_1type_crowd}).
We find excellent agreement between the $F_\alpha$ predicted by the DLEs~(\ref{eq:cd_DLLE}), the steady-state particle distributions implied by Eq.~(\ref{eq:cd_relation}) with Eq.~(\ref{eq:con_fd}), and the corresponding $F_\alpha$ obtained by averaging over KMC simulations of the ME~(\ref{df_ME}) with Eq.~(\ref{eq:W_diff2}).
As predicted by Eq.~(\ref{eq:cd_relation}) with Eq.~(\ref{eq:con_fd}), and
as in the case of free diffusion, we find that the $F_\alpha^\textrm{(s.s.)}$
in Fig.~\ref{fig:3doman_1type_crowd} are independent of the system geometry considered. Furthermore, as in the case of free diffusion, we find that the steady-state particle distributions in Fig.~\ref{fig:3doman_1type_crowd} are spatially inhomogeneous, provided that we do not have $N_i = 1$ for all $i$. Comparison of Figs.~\ref{fig:3doman_1type_free} and~\ref{fig:3doman_1type_crowd} shows that crowding tends to reduce spatial inhomogeneity in the steady-state particle concentration. Furthermore, comparison of Figs.~\ref{fig:3doman_1type_free} and~\ref{fig:3doman_1type_crowd} shows that crowding reduces the overshoot in $F_1$ in Fig.~\ref{fig:3doman_1type_free}(b). Indeed, decreasing the effects of crowding in Fig.~\ref{fig:3doman_1type_crowd}(b) by decreasing the value of $\langle N_i \rangle$ in the system we obtain, upon repeating the KMC simulations in Fig.~\ref{fig:3doman_1type_crowd}(b), an overshoot in~$F_1$.

\begin{figure}[t!]
\includegraphics[width=\columnwidth]{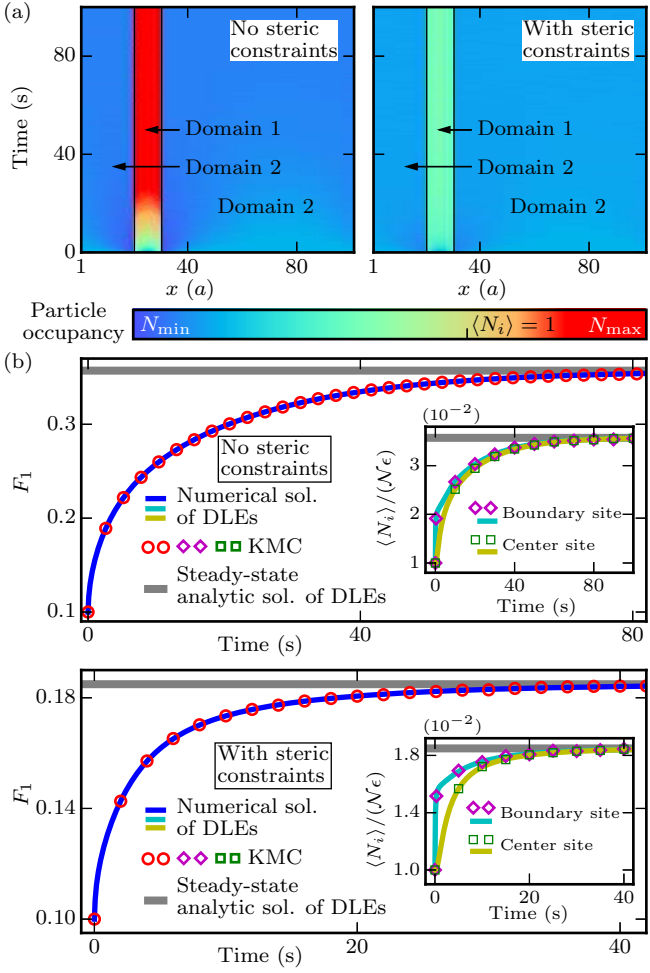} 
\caption{Single-species diffusion in inhomogeneous media for $d=1$ and two domains with $1/\tau_1=16$~s$^{-1}$ and $1/\tau_2=80$~s$^{-1}$. We used $K=100$ with $M_1=10$, periodic boundary conditions, and a homogeneous initial particle occupancy $N_i(0)=40\epsilon$ with $\epsilon=1/100$. (a) Average particle occupancy $\langle N_i \rangle$ as a function of time obtained from KMC simulations of the ME~(\ref{df_ME}) with Eq.~(\ref{eq:W_diff}) and Eq.~(\ref{eq:W_diff2}) for free diffusion (left panel) and diffusion under steric constraints (right panel), respectively. The black vertical lines show the domain boundaries. The minima and maxima of the plotted $\langle N_i \rangle$ are $(N_{\rm{min}},N_{\rm{max}})=(0.1,1.4)$, and we use the same color bar in the left and right panels. (b) Temporal evolution of the average fraction of all particles located in domain 1, $F_1$, for free diffusion (upper panel) and diffusion under steric constraints (lower panel). The insets show the temporal evolution of $\langle N_i \rangle$ and
$\phi_i$ for individual lattice sites at the center and boundary of domain 1. The dashed curves indicate numerical solutions of the DLEs~(\ref{eq:df_DLLE}) or~(\ref{eq:cd_DLLE}), the symbols denote averages over KMC simulations of the ME~(\ref{df_ME}) with Eq.~(\ref{eq:W_diff}) or Eq.~(\ref{eq:W_diff2}), and the gray lines show steady-state analytic solutions obtained from Eq.~(\ref{eq:df_sol2}) or Eq.~(\ref{eq:cd_relation}) with Eq.~(\ref{eq:con_fd}). All KMC results were averaged over 1000 independent realizations each.}
\label{fig:1D}
\end{figure}

\begin{figure}[t!]
\includegraphics[width=\columnwidth]{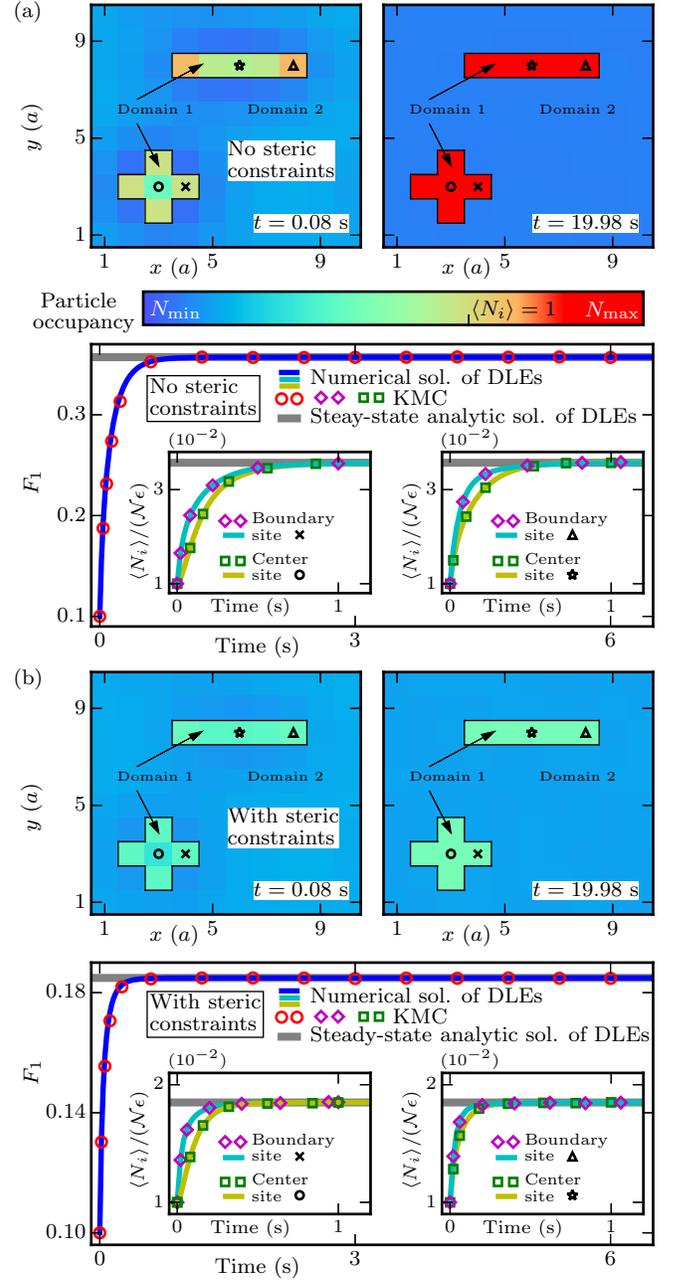}
\caption{Same results as in Fig.~\ref{fig:1D}, but for $d=2$ with domain~1 split up into two sub-domains, for (a) free diffusion and (b) diffusion
under steric constraints. The insets in the lower panels show the temporal evolution of $\langle N_i \rangle$ and $\phi_i$ for the individual lattice sites at the center and boundary of domain~1 indicated in the upper panels. As in Fig.~\ref{fig:1D}, all results were obtained from the DLEs~(\ref{eq:df_DLLE}) or~(\ref{eq:cd_DLLE}), the steady-state relations in Eq.~(\ref{eq:df_sol2}) or Eq.~(\ref{eq:cd_relation}) with Eq.~(\ref{eq:con_fd}), and KMC simulations of the ME~(\ref{df_ME}) with Eq.~(\ref{eq:W_diff}) or Eq.~(\ref{eq:W_diff2}).
We use the same color bar as in Fig.~\ref{fig:1D}.}
\label{fig:2D_1type}
\end{figure}

Figures~\ref{fig:1D} and~\ref{fig:2D_1type} provide detailed comparisons
between diffusion in inhomogeneous media for $d=1$ and $d=2$, for free diffusion
as well as diffusion under steric constraints. For all the scenarios considered
in Figs.~\ref{fig:1D} and~\ref{fig:2D_1type} we obtain excellent agreement between the average system properties predicted by the DLEs~(\ref{eq:df_DLLE}) or~(\ref{eq:cd_DLLE}), the steady-state relations in Eq.~(\ref{eq:df_sol2}) or Eq.~(\ref{eq:cd_relation}) with Eq.~(\ref{eq:con_fd}), and KMC simulations of the ME~(\ref{df_ME}) with Eq.~(\ref{eq:W_diff}) or Eq.~(\ref{eq:W_diff2}).
We first consider a system with $d=1$ and two distinct domains, with $M_1=K/10$ and $\tau_1=5\,\tau_2$ (see Fig.~{\ref{fig:1D}}). Starting from homogeneous initial conditions in $N_i$, we find a net flux of particles from domain 2 into domain 1, until the system reaches its steady state. In particular, domain 1 ``fills up'' from its boundaries inwards [see Fig.~\ref{fig:1D}(a)]. To quantify these observations we calculated, in addition to $F_1$, the average occupation number of individual lattice sites located at the center and at the boundary of domain 1, as a function of time [see Fig.~\ref{fig:1D}(b)]. We indeed find that the $\langle N_i \rangle$ for center cites in domain 1 lag behind the $\langle N_i \rangle$ for boundary sites in domain 1 in
their approach towards the steady state. Finally, we note that, compared to free diffusion, steric constraints produce a more rapid approach towards the steady state of the system in Fig.~\ref{fig:1D}.

In Fig.~\ref{fig:2D_1type} we consider diffusion in inhomogeneous media with the same system parameter values as in Fig.~\ref{fig:1D}, but for $d=2$ rather
than $d=1$ with domain~1 split up into two sub-domains. As predicted by the
steady-state analytic solution in Eq.~(\ref{eq:df_sol2}) for free diffusion
and by Eq.~(\ref{eq:cd_relation}) with Eq.~(\ref{eq:con_fd}) for diffusion under steric constraints, we find identical $F_{1,2}^\textrm{(s.s.)}$ for $d=1$ and $d=2$ in Figs.~\ref{fig:1D} and~\ref{fig:2D_1type}, for free diffusion as well as diffusion under steric constraints. However, for $d=2$ the system approaches its steady state more rapidly than for $d=1$. This can be understood by noting that, for the system geometries considered here, the length of the boundary separating domains 1 and 2 is larger for $d=2$ than for $d=1$, which is expected to facilitate particle exchange between distinct domains. Consistent with our results for $d=1$ in Fig.~\ref{fig:1D}, the evolution of the $d=2$ system in Fig.~\ref{fig:2D_1type} towards its steady state is more rapid for diffusion under steric constraints than for free diffusion, and the $\langle N_i \rangle$ for center cites in domain 1 lag behind the $\langle N_i \rangle$ for boundary sites in domain 1 in their approach towards the steady state.

\begin{figure}[t!]
\includegraphics[width=\columnwidth]{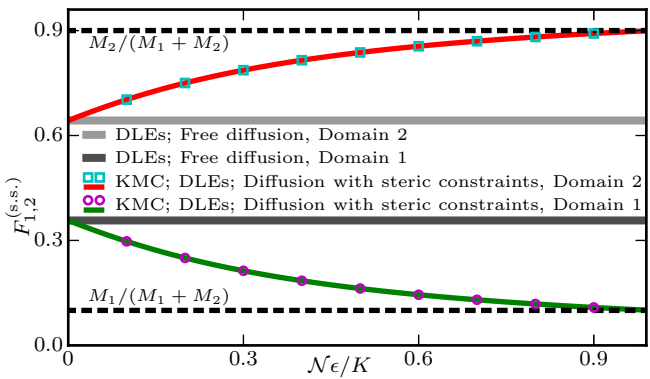}
\caption{$F_{1,2}^\text{(s.s.)}$ versus average particle occupancy per lattice site in the system, $\mathcal{N}\epsilon/K$, for the system in Fig.~\ref{fig:2D_1type}. As in Fig.~\ref{fig:2D_1type}, all results were obtained from the steady-state relations in Eq.~(\ref{eq:df_sol2}) and Eq.~(\ref{eq:cd_relation}) with Eq.~(\ref{eq:con_fd})
(horizontal solid lines and solid curves, respectively), and KMC simulations of the ME~(\ref{df_ME}) with Eq.~(\ref{eq:W_diff2}) (symbols).}
\label{fig:Frac_vs_avgDen}
\end{figure}

To further quantify the role of steric constraints in diffusion in inhomogeneous media we calculated, for the system in Fig.~\ref{fig:2D_1type}, the dependence of $F_{1,2}^\textrm{(s.s.)}$ on the (normalized) particle number in the system
(average particle occupancy per lattice in the system) $\mathcal{N} \epsilon/K$
for particles diffusing under steric constraints (see Fig.~\ref{fig:Frac_vs_avgDen}). Again, we obtain excellent agreement between the average system properties predicted by the steady-state particle distribution in Eq.~(\ref{eq:cd_relation}) with Eq.~(\ref{eq:con_fd}) and the corresponding results obtained from KMC simulations of the ME~(\ref{df_ME}) with Eq.~(\ref{eq:W_diff2}). We find that, as $\mathcal{N} \epsilon/K \to 0$, steric constraints become increasingly irrelevant and our results for $F_{1,2}^\textrm{(s.s.)}$ approach the corresponding results for free diffusion in Eq.~(\ref{eq:df_sol2}), with $F_2^\textrm{(s.s.)}/F_1^\textrm{(s.s.)} \to M_2 \tau_2/M_1 \tau_1$ as $\mathcal{N} \epsilon/K \to 0$. In contrast, as $\mathcal{N} \epsilon/K$ is increased, the effects of steric constraints become more and more pronounced, with $F_{2}^\textrm{(s.s.)}/F_1^\textrm{(s.s.)} \to M_{2}/M_1$ as $\mathcal{N} \epsilon/K \to 1$. Thus, depending on the values of $M_2/M_1$ and $\tau_2/\tau_1$ considered, steric constraints can tend to increase or decrease the inhomogeneity in $F_\alpha^\textrm{(s.s.)}$.

\subsection{Multiple particle species}
\label{subsec:multi-type}

In the presence of steric constraints, the diffusion of one particle species
can be affected \cite{Fanelli2010,Fanelli2013,Kahraman2016,Li2017} by the diffusion of other particle species in the system, and vice versa. We generalize here our formalism to allow for $S$ distinct particle species. We denote the occupation number at lattice site $i$ associated with particle species $s=1,2,\dots S$ by $N_{i;s}$, with $N_{i;s}/\epsilon_s$ corresponding to the number of particles of species $s$ at lattice site $i$ so that
\begin{equation}
0 \leq \sum_{s=1}^S N_{i;s} \leq 1
\end{equation}
for all $i$. Furthermore, we denote the hopping rate of particle species $s$ in domain $\alpha$ by $1/\tau_{\alpha(i);s}$. We assume that the total number of particles of each species in the system is conserved:
\begin{align} \label{eq:prob_conserv3}
\frac{1}{\mathcal{N}_s \epsilon_s} \sum \limits_{i=1}^{K} N_{i;s}= 1 \,,
\end{align}
where $\mathcal{N}_s$ is the total number of particles of species $s$
in the system. Redefining the lattice occupancy as ${\bf N}(t)=\{N_{i;s}(t)\}$
with $1\leq i \leq K$ and $1\leq s \leq S$, our stochastic lattice model of the diffusion of multiple particle species in inhomogeneous media under steric constraints is defined by the ME~(\ref{df_ME}) with the transition rate
\begin{align} \label{eq:W_diff3}
&W({\bf N};{\bf m}) = \frac{1}{2d}\sum\limits_{i,s} \frac{N_{i;s}}{\tau_{\alpha(i);s}\epsilon_s}
\delta(m_{i;s}+\epsilon_s)  \nonumber \\ & \times \sum\limits_{j\text{~n.n.~of~}i} \left(1-\sum_{l=1}^S N_{j;l}\right) \delta(m_{j;s}-\epsilon_s) \prod_{k\neq i,j} \delta(m_{k;s}) \,,
\end{align}
where the array of jumps in lattice occupancy ${\bf m}=\{m_{i;s}\}$ with
$1\leq i \leq K$ and $1\leq s \leq S$. Note that the factor $\left(1-\sum_{l=1}^S N_{j;l}\right)$ in the above transition rate couples the lattice occupancies
associated with distinct particle species.

Denoting the average occupation number of particle species $s$ at lattice site $i$ by $\phi_{i;s}(t)=\langle N_{i;s}(t) \rangle$ and proceeding as in Sec.~\ref{secFreeDiff}, we find that the DLEs associated with the ME~(\ref{df_ME}) with Eq.~(\ref{eq:W_diff3}) are given by 
\begin{align} \nonumber
\frac{d \phi_{i;s}}{d t} =& -\frac{\phi_{i;s}}{2d\tau_{\alpha(i);s}} \sum_{j\text{~n.n.~of~}i}
\left(1-\sum_{l=1}^S \phi_{j;l}\right)
\\\label{eq:cd_DLLE2}&
+\frac{1-\sum_{l=1}^S \phi_{i;l}}{2d}\sum_{j\text{~n.n.~of~}i}\frac{\phi_{j;s}}{\tau_{\alpha(j);s}} \,,
\end{align}
where, similarly as in Secs.~\ref{secFreeDiff} and~\ref{subsec:one-type}, the negative (positive) terms correspond to the hopping of particles of species~$s$ away from (to) lattice site $i$. As in Secs.~\ref{secFreeDiff} and~\ref{subsec:one-type}, the DLEs~(\ref{eq:cd_DLLE2}) are, for a given set of initial conditions $\{\phi_{i;s}(0)\}$, amenable to direct numerical solution using standard methods \cite{Mathematica}, which uniquely specifies $\phi_{i;s}(t)$ for all $i$, all $s$, and all $t$. From the $\phi_{i;s}(t)$
the average fraction of all particles of species $s$ in domain $\alpha$, $F_{\alpha;s}$, can be computed by evaluating
\begin{equation} \label{FalphadefM}
F_{\alpha;s}=\frac{\sum_{i\textrm{~in~domain~}\alpha}\phi_{i;s}}{\sum_{i=1}^K
\phi_{i;s}}\,.
\end{equation}

To calculate the steady-state particle distribution $F_{\alpha;s}^{\textrm{(s.s.)}}$
associated with Eq.~(\ref{FalphadefM}) we follow steps analogous to those in Sec.~\ref{subsec:one-type}. Equation~(\ref{eq:cd_DLLE2}) suggests that, in the steady state of the system, all the $\phi_{i;s}$ lying in a particular domain $\alpha$ with hopping rate $1/\tau_{\alpha;s}$ take the same value $\phi^{(\alpha;s)}$. In the steady state of the system, Eq.~(\ref{eq:cd_relation}) then generalizes to
\begin{align} \label{eq:cd_relationM}
\frac{1}{1-\sum_{l=1}^S\phi^{(\alpha;l)}} \frac{\phi^{(\alpha;s)}}{\tau_{\alpha;s}} = \frac{1}{1-\sum_{l=1}^S\phi^{(\beta;l)}} \frac{\phi^{(\beta;s)}}{\tau_{\beta;s}}\,.
\end{align}
Together with Eq.~(\ref{eq:prob_conserv3}), Eq.~(\ref{eq:cd_relationM}) allows (numerical) calculation of $\phi^{(\alpha;s)}$ for each domain and each particle
species. The resulting solutions for the steady-state particle distribution are expected to be unique. For instance, consider a system with only two domains $\alpha$ and $\beta$. Upon applying Eq.~(\ref{eq:cd_relationM}) to
the two particle species $s$ and $k$ and dividing the resultant relations,
we find
\begin{align} \label{eq:cd_relationMr}
\frac{\phi^{(\alpha;s)}}{\tau_{\alpha;s}} \frac{\tau_{\alpha;k}}{\phi^{(\alpha;k)}}= \frac{\phi^{(\beta;s)}}{\tau_{\beta;s}} \frac{\tau_{\beta;k}}{\phi^{(\beta;k)}}\,.
\end{align}
If $\phi^{(\alpha;s)}$ is changed from a steady-state solution $\phi^{(\alpha;s)}=\phi^{(\alpha;s)}_+$ to a steady-state solution $\phi^{(\alpha;s)}=\phi^{(\alpha;s)}_-$, with $\phi^{(\alpha;s)}_+ > \phi^{(\alpha;s)}_-$ and $0 \leq \phi^{(\alpha;s)}_\pm \leq1$, Eq.~(\ref{eq:prob_conserv3}) requires a corresponding increase in $\phi^{(\beta;s)}$. According to Eq.~(\ref{eq:cd_relationMr}), such a change in the distribution of particle species $s$ requires a decrease in $\phi^{(\alpha;k)}/\phi^{(\beta;k)}$ which, because of Eq.~(\ref{eq:prob_conserv3}), can only be achieved if $\phi^{(\alpha;k)}$ decreases and $\phi^{(\beta;k)}$ increases, thus violating Eq.~(\ref{eq:cd_relationM}).

The analogy with the thermodynamic formalism describing the self-assembly of particle aggregates in dilute solutions \cite{BenShaul1994,Safran2003} drawn in Sec.~\ref{secDLEfreePD} for free diffusion can be extended to include steric constraints as well as multiple diffusing particle species. Equation~(\ref{eq:prob_conserv3}) mandates that the particle number is conserved for each particle species, yielding a distinct $\mu_s$ for each particle species $s$. As in Sec.~\ref{secDLEfreePD}, Eq.~(\ref{eq:cd_relationM}) can then be rewritten as
\begin{align} \label{eq:thermo_multi1}
\frac{\phi^{(\alpha;s)}}{1-\sum_{l=1}^S \phi^{(\alpha;l)}}= e^{\mu_s-\epsilon_{\alpha;s}} \, ,
\end{align}
where $\epsilon_{\alpha;s}=\log \left(\tau_0/\tau_{\alpha;s} \right)$ and $\tau_0$ is a constant. Equation~(\ref{eq:thermo_multi1}) implies that
\begin{align}
\frac{\sum_{s=1}^S \phi^{(\alpha;s)}}{1-\sum_{s=1}^S \phi^{(\alpha;s)}}= \sum_{s=1}^S e^{\mu_s-\epsilon_{\alpha;s}} \, ,
\end{align}
which can be rearranged to  
\begin{align} \label{eq:thermo_multi2}
 1-\sum_{s=1}^S\phi^{(\alpha;s)} = \frac{1}{1+\sum_{s=1}^S e^{\mu_s-\epsilon_{\alpha;s}}} \, .
\end{align}
Insertion of Eq.~(\ref{eq:thermo_multi2}) back into Eq.~(\ref{eq:thermo_multi1})
yields the steady-state distribution of particles diffusing through inhomogeneous media under steric constraints,
\begin{align} \label{eq:thermo_multi3}
 \phi^{(\alpha;s)} = \frac{e^{\mu_s-\epsilon_{\alpha;s}}}{1+\sum_{l=1}^S e^{\mu_l-\epsilon_{\alpha;l}}} \, ,
\end{align}
where the $\mu_s$ are determined by Eq.~(\ref{eq:prob_conserv3}) via
\begin{align} \label{eq:thermo_multi4}
 \sum_{\beta=1}^D \frac{M_\beta e^{\mu_s-\epsilon_{\beta;s}}}{1+\sum_{l=1}^S e^{\mu_l-\epsilon_{\beta;l}}}  = \mathcal{N}_s\epsilon_s \, ,
\end{align}
which couples distinct domains and particle species. Equations~(\ref{eq:thermo_multi3}) and (\ref{eq:thermo_multi4}) reduce the calculation of $\phi^{(\alpha;s)}$ to the solution of Eq.~(\ref{eq:thermo_multi4}). The special case $S=1$ in
Eqs.~(\ref{eq:thermo_multi3}) and (\ref{eq:thermo_multi4}) yields the steady-state
distribution of particles diffusing through inhomogeneous media under steric constraints for a single diffusing particle species (see Sec.~\ref{subsec:one-type}).

\subsection{Simulation of multi-species diffusion under steric constraints}
\label{secSimMulti}

\begin{figure}[t!]
\includegraphics[width=\columnwidth]{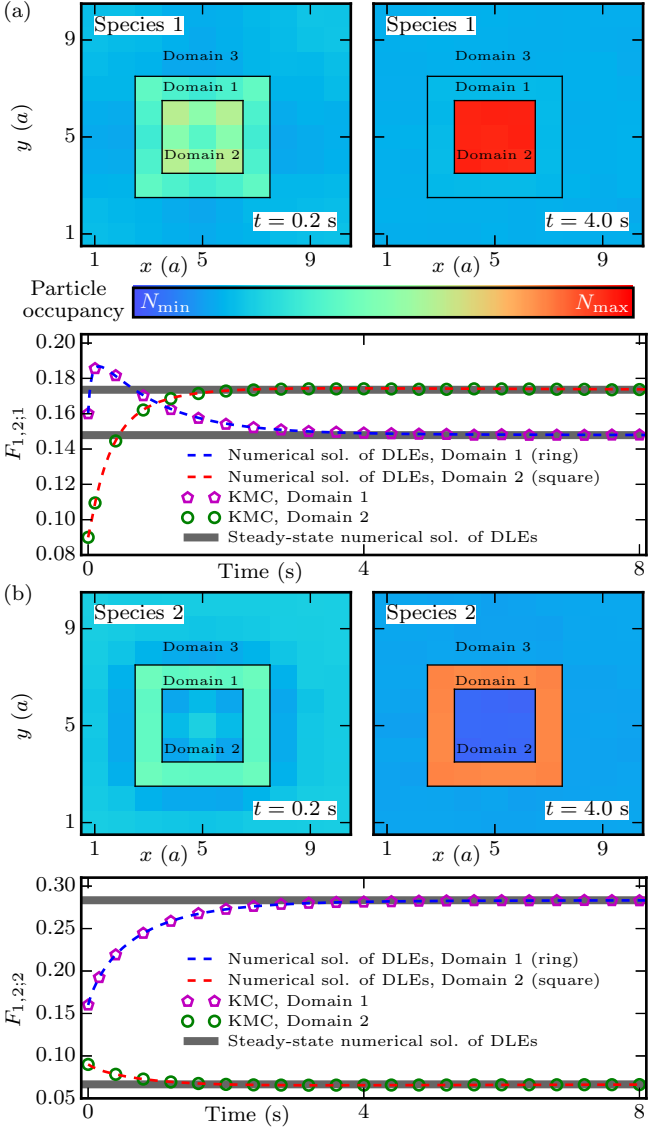}
\caption{Two-species diffusion under steric constraints for $d=2$ and three domains with $\left(1/\tau_{1;1},1/\tau_{2;1},1/\tau_{3;1}\right)=\left(32,16,80\right)$~s$^{-1}$
and $\left(1/\tau_{1;2},1/\tau_{2;2},1/\tau_{3;2}\right)=\left(3.2,8,16\right)$~s$^{-1}$
employing the same system geometry as in Figs.~\ref{fig:3doman_1type_free} and~\ref{fig:3doman_1type_crowd} for (a) species 1 and (b) species 2. We used $K=100$ with $M_1=16$ and $M_2=9$, periodic boundary conditions, and the homogeneous initial particle distributions $N_{i;1}(0)=30\epsilon_1$ and $N_{i;2}(0)=30\epsilon_2$ with $\epsilon_1=\epsilon_2=1/100$. The upper panels in (a) and (b) show $\langle N_{i;1} \rangle$ and $\langle N_{i;2} \rangle$ at the indicated $t$ obtained from KMC simulations of the ME~(\ref{df_ME}) with Eq.~(\ref{eq:W_diff3}). The minima and maxima of the $\langle N_{i;s} \rangle$ in (a) and (b) are $(N_{\rm{min}},N_{\rm{max}})=(0.2,0.6)$. The lower panels in (a) and (b) show the temporal evolution of $F_{\alpha;s}$ for $\alpha=1,2$ and $s=1,2$. The dashed curves indicate numerical solutions of the DLEs~(\ref{eq:cd_DLLE2}), the symbols denote averages over KMC simulations of the ME~(\ref{df_ME}) with Eq.~(\ref{eq:W_diff3}), and the gray lines show the $F_{\alpha;s}^\textrm{(s.s.)}$ obtained from Eq.~(\ref{eq:cd_relationM}) with Eq.~(\ref{eq:prob_conserv3}). The KMC results were averaged over 1000 independent realizations.}
\label{fig:3doman_2type_crowd}
\end{figure}

\begin{figure}[t!]
\includegraphics[width=\columnwidth]{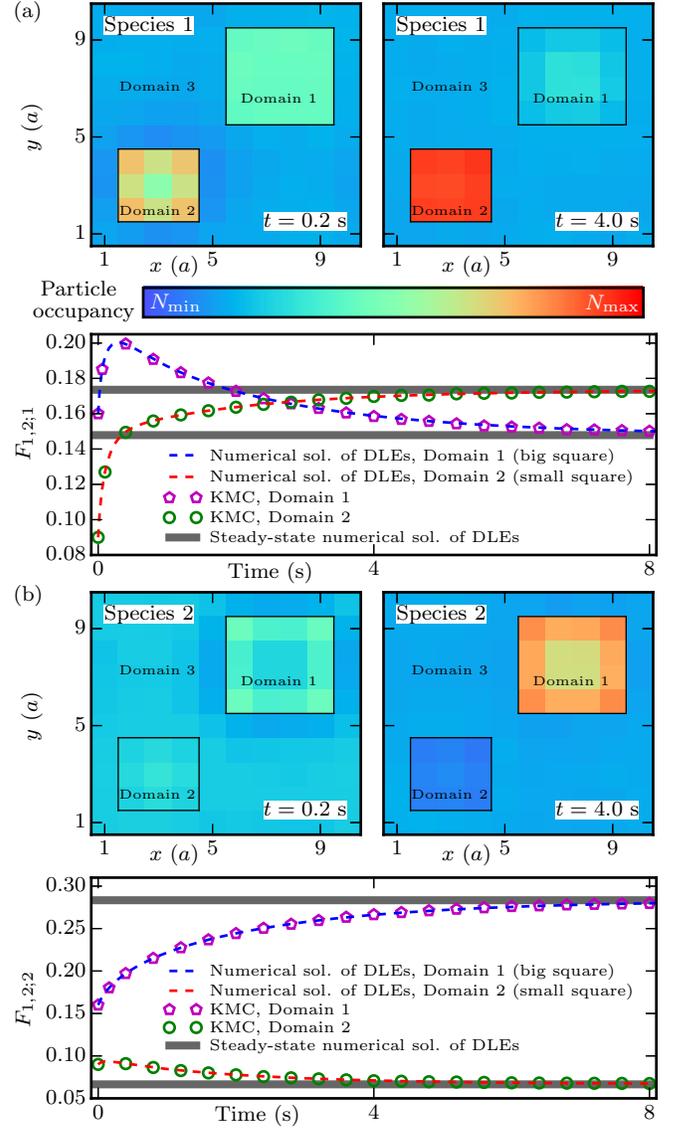}
\caption{Same results as in Fig.~\ref{fig:3doman_2type_crowd}, but for the system geometry shown in the upper panels of (a) and (b) using the same values of $K$ and $M_{1,2,3}$ as in Fig.~\ref{fig:3doman_2type_crowd}. As in Fig.~\ref{fig:3doman_2type_crowd}, all results were obtained from the DLEs~(\ref{eq:cd_DLLE2}), the ME~(\ref{df_ME}) with Eq.~(\ref{eq:W_diff3}), and the steady-state relations in Eq.~(\ref{eq:cd_relationM}) with Eq.~(\ref{eq:prob_conserv3}).
We use the same color bar as in Fig.~\ref{fig:3doman_2type_crowd}.}
\label{fig:3doman_2type_crowd2}
\end{figure}

As in Secs.~\ref{secSimFD} and~\ref{secSimSingleSteric}, we tested the accuracy of the particle distributions predicted by the DLEs~(\ref{eq:cd_DLLE2}), with the steady-state particle distributions implied by Eq.~(\ref{eq:cd_relationM}) with Eq.~(\ref{eq:prob_conserv3}), by carrying out KMC simulations of the ME~(\ref{df_ME}) with Eq.~(\ref{eq:W_diff3}) using the Next Subvolume Method \cite{Elf2003} (see Figs.~\ref{fig:3doman_2type_crowd} and~\ref{fig:3doman_2type_crowd2}). We considered the same system geometries and parameter values as in Figs.~\ref{fig:3doman_1type_free} and~\ref{fig:3doman_1type_crowd}, but allowed for a second particle species
with hopping rates that were reduced compared to the hopping rates of particle
species 1. We find excellent agreement between the $F_{\alpha;s}$ predicted by the DLEs~(\ref{eq:cd_DLLE2}), the steady-state particle distributions implied by Eq.~(\ref{eq:cd_relationM}) with Eq.~(\ref{eq:prob_conserv3}), and the corresponding $F_{\alpha;s}$ obtained by averaging over KMC simulations of the ME~(\ref{df_ME}) with Eq.~(\ref{eq:W_diff3}). As predicted by Eq.~(\ref{eq:cd_relationM}) with Eq.~(\ref{eq:prob_conserv3}), we obtain spatially inhomogeneous steady-state particle distributions, with $F_{\alpha;s}^\textrm{(s.s.)}$ being independent of the arrangement and shape of domains. Comparison of Figs.~\ref{fig:3doman_2type_crowd} and~\ref{fig:3doman_2type_crowd2} with Fig.~\ref{fig:3doman_1type_crowd} shows that the presence of more than one diffusing particle species can have complex effects on the temporal evolution of $F_{\alpha;s}$. In particular, Figs.~\ref{fig:3doman_2type_crowd} and~\ref{fig:3doman_2type_crowd2} show that interactions between diffusing particle species via steric constraints
can slow down the approach towards the steady state of the system, and alter
even basic qualitative features of the temporal evolution of $F_{\alpha;s}$. For instance, we find a pronounced overshoot in $F_{1;1}$ in Fig.~\ref{fig:3doman_2type_crowd} as well as Fig.~\ref{fig:3doman_2type_crowd2}, while no such overshoot occurs for $F_1$ in Fig.~\ref{fig:3doman_1type_crowd}.

\section{Summary and conclusions}
\label{SumCon}

Diffusion can be conceptualized \cite{Chandrasekhar1943,Berg1993,Codling2008} as the random hopping of particles between neighboring lattice sites with, in the case of diffusion in inhomogeneous media, distinct particle hopping rates in distinct spatial domains in the system. Starting from the MEs \cite{Gardiner1985,Kampen1992} describing the random hopping of particles in inhomogeneous media, we have
derived here the DLEs governing diffusion in inhomogeneous media in arbitrary
spatial dimensions for free diffusion as well as diffusion under steric constraints. For a given initial particle distribution, the DLEs can be readily solved numerically. We have also obtained general analytic expressions for the steady-state particle distributions for free diffusion and, in special cases, diffusion under steric constraints in inhomogeneous media. We find that the particle distributions obtained from the DLEs are, for both transient and asymptotic regimes, in excellent agreement with averages over KMC simulations of the underlying MEs. We used here $\epsilon=1/100$, which is suitable \cite{Kahraman2016,Li2017} for modeling protein diffusion in cell membranes. For general $\epsilon$, the MEs and DLEs are expected to yield similarly good agreement if $\epsilon \lessapprox 1/10$ \cite{Li2017}. The origin of the observed agreement between MEs and DLEs may lie \cite{Li2017} in the conservation of particle number in the stochastic lattice models of diffusion considered here, which constrains the fluctuations in the MEs \cite{Haselwandter2002}.

From a computational perspective, solution of the DLEs obtained here for a system composed of $S$ distinct particle species hopping between $K$ lattice sites amounts to the solution of $S\times K$ coupled, first-order ordinary differential equations, which can be efficiently achieved, starting from
a given set of initial conditions, using standard methods \cite{Mathematica}. If only the average steady state of the system is of interest, the computational complexity of the problem can be reduced further with, for a system containing $D$ domains with distinct particle hopping rates, solution of only $S\times D$ algebraic equations being required to predict the steady-state particle distribution. Note, in particular, that the number of DLEs associated with a given system is independent of the number of particles in the system. Thus, the DLEs provide a particularly favorable approach for situations in which the particle number is large, which is often the case when modeling
experiments on diffusion in inhomogeneous media 
\cite{Siggia2000,Czondor2012,Cantrell1999,Fagan1999,Cronin2003,Delay2005,Hoteit2002,LaBolle2000,Zhan2009,Fieremans2010,Marcowith2010,Zhang2000}.

We find that the average fraction of particles in a given domain may show---depending on key system properties such as the system geometry, the initial conditions used, the dimensionality of the system, and the number of distinct diffusing particle species---a complex approach towards the steady state of the system. For instance, depending on the detailed system properties, the average fraction of particles in a given domain may overshoot when approaching the steady state of the system, due to a slow equilibration between domains with distinct particle hopping rates. We find that the magnitude of this overshoot depends critically on the separation of domains, with larger domain separations yielding a more pronounced overshoot. For systems comprising only a single particle species, molecular crowding tends to reduce the magnitude of the overshoot in the average fraction of particles in a given domain. Our results suggest that, in systems with many domains with distinct particle hopping rates, the particle distribution can show a highly non-monotonic temporal evolution towards the steady state of the system, with a hierarchy of timescales set by the particle hopping rates in different domains and the system geometry. We find that the complexity of the temporal evolution of the particle distribution can be further increased if the system comprises multiple particle species interacting via steric constraints. In this case, the presence of one particle species can, for instance, induce an overshoot in the average fraction of another particle species in a given domain. Furthermore, we find that interactions between multiple particle species via steric constraints can slow down the approach of the particle distribution towards the steady state of the system.

The DLEs derived here suggest that, in the steady state of the system, the average fraction of particles in a given domain is independent of most system properties, even if the particles interact via steric constraints. We find that the average steady-state concentration of particles is uniform in each domain, and only depends on the number of lattice sites in each domain, the particle hopping rates, the number of distinct particle species in the system, and the total number of particles of each particle species in the system. In particular, the DLEs derived here suggest that the average steady-state concentration of particles in each domain is independent of the arrangement and shape of domains. While we have focused here on the deterministic parts of the lattice Langevin equations associated with diffusion in inhomogeneous media, the formalism employed here can be extended \cite{McKane2004,Haselwandter2007,Haselwandter2007PRL,Haselwandter2007EPL,Lugo2008,Butler2009,Butler2011,Biancalani2017} to carry out a systematic analysis of the fluctuations induced by the random
hopping of particles in inhomogeneous media, and to connect the DLEs derived
here to generalized diffusion equations with spatially-varying diffusion coefficients \cite{Kampen1987,Kampen1988,Miyazawa1987,Kampen1992,Haselwandter2011,Haselwandter2015,Li2017}.

The general mathematical results obtained in this article are of relevance to diffusion in inhomogeneous media in a variety of different experimental systems
\cite{Fieremans2010,Siggia2000,Czondor2012,Cantrell1999,Fagan1999,Cronin2003,Delay2005,Hoteit2002,LaBolle2000,Zhan2009,Zhang2000,Marcowith2010}.
An important point here is that in complex, heterogeneous systems it is often
not clear from the outset whether a simple random walk model with
spatially varying hopping rates can capture the basic features of the particle
dynamics. Our results show that the steady-state distributions of particles in inhomogeneous media may be used to deduce key features of the particle dynamics even if detailed system properties, such as the shape and arrangement of distinct domains in the system, are not known. For instance, synaptic receptors diffuse randomly through cell membranes with hopping rates that are reduced inside synaptic membrane domains \cite{Czondor2012}. In addition to diffusion, however, synaptic receptors may show complex interactions with other molecules in the cell membrane, and undergo recycling via endo- and exocytosis \cite{Czondor2012}. For a given set of experimental conditions, our results could be used, for instance, to formulate tests of whether such additional processes substantially affect the measured steady-state receptor distribution in the membrane, or whether the measured steady-state
receptor distribution is primarily set by the observed inhomogeneity in the receptor hopping rates. For general experimental realizations of diffusion in inhomogeneous media \cite{Fieremans2010,Siggia2000,Czondor2012,Cantrell1999,Fagan1999,Cronin2003,Delay2005,Hoteit2002,LaBolle2000,Zhan2009,Zhang2000,Marcowith2010},
the DLEs and corresponding analytic results obtained here may similarly be employed to ascertain whether spatially inhomogeneous particle hopping rates are already sufficient to explain a particular, spatially inhomogeneous particle distribution observed in experiments, or whether more complicated microscopic mechanisms and interactions must be invoked in order to understand experimental data on the distribution of randomly diffusing particles in inhomogeneous media.

\acknowledgments{We thank F. Pinaud for helpful discussions on protein diffusion
in membranes. This work was supported by NSF award number DMR-1554716, an Alfred P. Sloan Research Fellowship in Physics, the James H. Zumberge Faculty Research and Innovation Fund at USC, and the USC Center for High-Performance Computing.}


\end{document}